\documentclass[twocolumn, apj, iop, superscriptaddress,floatfix,showpacs]{aastex701}

\usepackage{graphicx}
\usepackage{amssymb}
\usepackage{amsmath}
\usepackage{booktabs}
\usepackage{multirow}

\begin{document}

\title{NuGNN: a Graph Neural Network for Nuclear Reaction Network Equations}

\author{C.~H.~\surname{Kim}}
\affiliation{Department of Physics, Sungkyunkwan University, Suwon 16419, Republic of Korea}
\email{chkim@phys.kim}

\author{K.~Y.~\surname{Chae}}
\affiliation{Department of Physics, Sungkyunkwan University, Suwon 16419, Republic of Korea}
\email{kchae@skku.edu}
\thanks{Fax: +82-31-290-7055}

\author{S.~\surname{Ko}}
\affiliation{Department of AI Systems Engineering, Sungkyunkwan University, Suwon 16419, Republic of Korea}
\email{shko0215@skku.edu}

\author{M.~R.~\surname{Mumpower}}
\affiliation{Obsidian Research, Fort Wayne, IN 46835, USA}
\affiliation{Department of Physics and Astronomy, University of Notre Dame, Notre Dame, IN, 46656, USA}
\email{matthew@mumpower.net}

\author{M.~S.~\surname{Smith}}
\affiliation{Stellar Science Solutions, Lenoir City, TN 37772, USA}
\email{Michael.Smith@stellarsciencesolutions.com}

\begin{abstract}

Nuclear reaction networks are a major computational bottleneck in astrophysical simulations when large isotope sets are required, because of the stiffness of the network equations and the repeated calls to Jacobian-based solvers required by implicit methods. In this work, we develop a deep learning surrogate solver for a large 690-isotope nuclear reaction network under general Type I X-ray burst conditions using a graph neural network, NuGNN. Unlike conventional fully connected or convolutional neural networks, NuGNN directly reflects the structure of the reaction network through heterogeneous isotope and reaction nodes and message-passing along reaction connections. The model is trained on data spanning many orders of magnitude in stellar temperature and density and in simulation time step size. We compare NuGNN with a Res-U-Net and fully connected neural network and find that NuGNN consistently achieves significantly better accuracy with errors of only a few percent. More importantly, when implemented in the network evolution code in place of the original solver, NuGNN successfully reproduces the final abundance patterns, whereas the other architectures fail to do so. We also show that the trained model can substantially improve computational speed, demonstrating its practical potential for large-scale simulations. These results show that graph neural networks provide a robust and promising framework for accurate surrogate modeling of large nuclear reaction networks. 

\end{abstract}

\section{\label{sec1}Introduction}

Nuclear reactions play an important role in many astrophysical environments, including Type I X-ray bursts, novae, core-collapse supernovae, and many others \cite[]{A.Arcones2017, H.Schatz2006, C.Kim2022, J.Cowan2021}. In these systems, nuclear reaction networks are solved over small time steps to describe the time evolution of isotopic abundances, thereby contributing to both the energy generation and the synthesis of heavy elements in the Universe. However, solving such networks is computationally demanding, especially when a large number of isotopes is involved \cite[]{F.Timmes1999, F.Timmes2000, A.Jermyn2023, G.Navo2023}. Astrophysical applications that accompany heavy nucleosynthesis often require tracking hundreds to thousands of isotopes. In conventional solvers, this leads to a rapidly increasing computational burden, as stable and converged solutions often require very small time steps together with repeated iterative updates of reaction rates and Jacobian matrices. In addition, the cost also grows substantially with the network size, since the number of involved reactions can increase up to $\sim$10$^{5}$ and the size of the Jacobian matrix scales as $N^2$, where $N$ is the number of isotopes in the network. Such costs become a serious bottleneck in large-scale simulations, when the network must be solved over many spatial zones and time steps. As a result, many practical simulations, particularly multi-dimensional ones, are forced to employ only reduced reaction networks that approximate the nuclear energy generation but do not track the abundances of the hundreds to thousands of isotopes altered in the thermonuclear burn \cite[]{F.Timmes2000, R.Farmer2016, E.Bravo2020, S.Bruenn2020}. 

Deep learning has become a powerful tool for modeling complex systems and has demonstrated strong performance in a wide range of applications \cite[]{A.Boehnlein2022}; it also shows great potential for nucleosynthesis studies \cite[]{M.Smith2024}. Recently, a few efforts have been made to replace the conventional solver with deep learning surrogates \cite[]{D.Fan2022, A.Grichener2025, X.Zhang2025}. However, achieving robust generalization with large nuclear reaction networks remains a major challenge. \cite{D.Fan2022} emulated the reaction step of 3-isotope network in MAESTROeX for the early stage of carbon fusion in Type Ia supernovae, training on data from a specific simulation with a fixed time step size. \cite{A.Grichener2025} extended this direction to 80- and 151-isotope networks for late burning stages in core-collapse supernova progenitors, where the training ranges of temperature and density spanned about 1--2 orders of magnitude and separate models were trained for different fixed time step sizes. \cite{X.Zhang2025} considered 3- and 13-isotope networks, with training ranges in temperature and density again spanning about 1--2 orders of magnitude, together with fixed time step sizes. We also note that these previous studies mainly relied on fully connected feed-forward architectures, including residual layers in \cite{D.Fan2022}.

\begin{figure*}[ht]
\centering
	\includegraphics[width=1\textwidth]{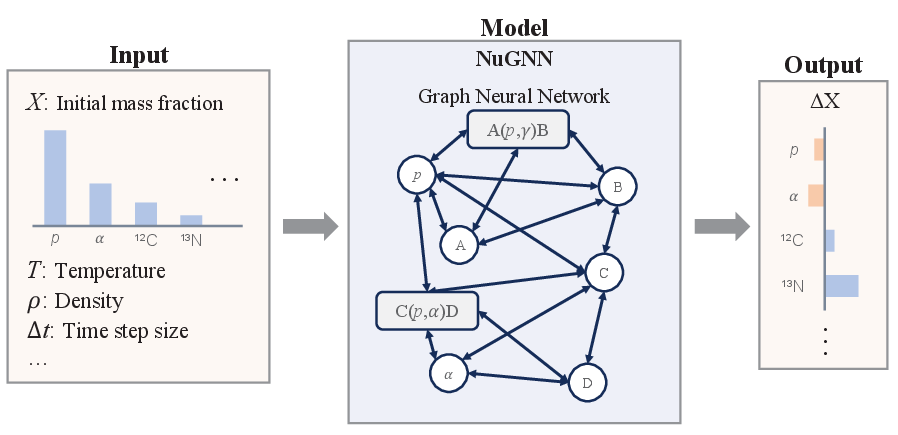}	 
\caption{\label{fig_flow}Schematic input--output overview of NuGNN for a single abundance update. NuGNN processes the input quantities using a graph neural network composed of isotope and reaction nodes and predicts the corresponding change in mass fraction, $\Delta X$. See the text for details.}
\end{figure*}

In this study, we utilized deep learning to develop a surrogate replacement for the traditional thermonuclear burn solver, specifically for a large 690-isotope nuclear reaction network under general conditions for Type I X-ray burst (XRB) simulations. We trained the model on data spanning many orders of magnitude in stellar temperature and density, as well as in simulation time step size. Rather than using conventional architectures such as fully connected or convolutional neural networks, we implemented a graph neural network that directly reflects the structure of the nuclear reaction network through isotope and reaction nodes and their connections. We demonstrated the robust performance of graph neural networks in this task through comparison with a U-Net based neural network and fully connected neural network. In the following sections, we describe the formulation of reaction network equations, data generation and preprocessing, neural network design, training procedure, and the test results, together with possible directions for further development. 


\section{\label{sec_nuc}Nuclear reaction network equations}

Nuclear reaction network equations describe the time evolution of the abundances of nuclides under nuclear reactions and decays \cite[]{WinNet}:
\begin{align}
\begin{split}
\dot{Y}_i &= f(Y_i; T, \rho) \\ 
 &= \sum_j N^i_j \lambda_j Y_j + \sum_{j,k} N^i_{j,k} \rho N_A \langle j,k \rangle Y_j Y_k \\
&+ \sum_{j,k,l} N^i_{j,k,l} \rho^2 N_A^2 \langle j,k,l \rangle Y_j Y_k Y_l, 
\label{eq_net}
\end{split}
\end{align}
where $Y_i$ is the abundance of species $i$, $T$ and $\rho$ are temperature and density, $N^i_j$, $N^i_{j,k}$, and $N^i_{j,k,l}$ represent the number of species $i$ destroyed or created in the reaction, $\lambda_j$, $\langle j,k \rangle$, and $\langle j,k,l \rangle$ are decay and reaction rates, and $N_A$ represents Avogadro's constant. In astrophysical environments, the set of coupled Equations~\ref{eq_net} for all species $i$, $j$, $k$, $l$ are repeatedly solved over small time steps to determine the time evolution of the system composition. The reaction network couples a large number of isotopes through reaction and decay channels including particle captures, photodisintegrations, beta decays, and so on. Therefore, the abundance evolution of each isotope depends not only on its own state but also on the abundances of many other isotopes connected to it through the reaction network. 

Nuclear reaction network equations are typically stiff, especially at high $T$ and $\rho$, where the solutions depend on a wide range of timescales \cite[]{W.Hix2006, C.Travaglio2013, A.Jermyn2023}. Because of this stiffness, explicit integration methods are usually impractical or unstable unless prohibitively small time steps are used. In practice, implicit solvers such as the backward Euler method are therefore widely adopted \cite[]{S.Bruenn2020, SkyNet, WinNet, PRISM}. These methods require the repeated solution of a time-linearized system involving the Jacobian matrix, which can become computationally expensive for large isotope sets \cite[]{W.Hix2006, C.Travaglio2013}.

A typical network solver at each time step first evaluates the reaction rates and Jacobian matrix using the current abundances, temperature, and density. The time step size is chosen based on convergence criteria and other physical or numerical constraints. These quantities are then used to solve Equation~\ref{eq_net} with an implicit integration method, typically through Newton-Raphson iterations. The resulting abundance change is added to the previous abundances to obtain the updated composition. In this work, we focus on replacing the most computationally demanding components, namely the Jacobian construction and implicit solver step, with a deep learning surrogate that directly predicts the abundance change for a given input state.

\section{\label{sec_data}Data Preparation}

In this section, we describe how we obtained the input-output pairs corresponding to one time step of the thermonuclear burn solver. Figure~\ref{fig_flow} illustrates the input--model--output workflow of NuGNN, in which the nuclear and thermodynamic quantities are processed by the graph neural network and mapped to the predicted abundance change. The inputs to our model included initial mass fraction $X$ (mass fraction at the current time), reaction rates, net flux term $f(Y; T, \rho)$ in Equation~\ref{eq_net} at the current time, separation energies, $T$, $\rho$, and simulation time step size $\Delta t$, where the Jacobian matrix is not needed as an input, because it can be reconstructed from these values. We used the mass fraction, $X=A\,Y$, instead of the abundance, $Y$, because the sum of the mass fractions is unity, providing a convenient check of mass conservation and solution convergence. The output (label) was the change in abundance $\Delta X$ after $\Delta t$ (see Section~\ref{sec_data_b} for details). We obtained these input-output pairs via executions of the reaction network code $\texttt{PRISM}$ \cite[]{PRISM} that uses the implicit Euler integration scheme with the Newton-Raphson method. 

\begin{table}[t]
\centering
\caption{\label{tab_range}Ranges of the input variables used in the training data. Each variable was either sampled within the range listed here or taken directly from the simulations. See the text for details.}
\begin{ruledtabular}
\begin{tabular}{ccc|ccc} 

& Physical quantity & & & Training data range & \\ \hline
& $X$ (Mass fraction) & & & [10$^{-20}$, 1] & \\
& $T$ (Temperature) & & & [10$^{6}$, 2.5$\times 10^{9}$] K & \\ 
& $\rho$ (Density) & & & [10$^{-2}$,  10$^{8}$] g cm$^{-3}$ & \\ 
& $\Delta t$ (Time step size) & && [10$^{-15}$, 10$^{-4}$] s & \\ 

\end{tabular}
\end{ruledtabular}
\end{table}


\subsection{\label{sec_data_a}Data Generation}

Table~\ref{tab_range} shows a summary of the training data ranges for $X$, $T$, $\rho$, and $\Delta t$. The ranges of $T$ and $\rho$ were determined to include general cases of XRB. We note that the use of variable $\Delta t$ enables the solver to be utilized under rapidly changing stellar conditions and adapt to various physical or numerical factors that can limit the time step size.

The training data were divided into three categories with increasing levels of randomness to improve the generalization of the model. Type A contains physically realistic samples taken directly from reaction network calculations along thermodynamic trajectories of XRB, where only $\Delta t$ is randomly sampled within the range. Type B contains augmented samples based on realistic compositions, where $X$ are randomly rescaled and $T$, $\rho$, and $\Delta t$ are randomly assigned. Type C contains fully random synthetic samples. This three-level construction was adopted to balance physical realism and broad input-space coverage, thereby improving model generalization. 

The thermodynamic trajectories were obtained from both X-ray burst simulations performed in this work and from the work of \cite{H.Schatz2001}. We performed XRB simulations with a 1-D stellar evolution code, Modules for Experiments in Stellar Astrophysics (MESA) \cite[]{B.Paxton2015}. The simulation setup followed that of \cite{Z.Meisel2018, C.Kim2022}, except that the accretion rate ($\dot{M}$) and metallicity ($Z$) were varied. We considered $\dot{M}$ = 0.07, 0.17, and 0.28 $\dot{M}_{\text{edd}}$, where $\dot{M}_{\text{edd}}$ represents the Eddington accretion rate, together with $Z$ = 0.01, 0.02, 0.05, and 0.10, where the hydrogen fraction ($X$) and helium fraction ($Y$) were varied according to $Y \approx 0.24 + 1.75Z$ and $X+Y+Z=1$ \cite[]{N.Lampe2016}. This yielded a total of 11 simulation cases, as the case with $\dot{M}$ = 0.07 $\dot{M}_{\text{edd}}$ with $Z$ = 0.10 did not successfully produce the bursts. Because MESA uses adaptive mesh refinement, the number and locations of zones vary during the simulation. We therefore extracted thermodynamic trajectories at nine different column depths rather than following fixed zone numbers. 

To obtain the Type A and Type B input datasets, the extracted trajectories were subsequently post processed in reaction network calculations---$\texttt{PRISM}$ for the trajectories from MESA and $\texttt{WinNet}$ \cite[]{WinNet} for the trajectory from \cite{H.Schatz2001}. The calculations were performed with $\beta^+$ decays, $(p,\gamma)$, $(\alpha,\gamma)$, $(p,\alpha)$, and inverse reaction of these, as XRBs mainly proceed with $\alpha p$- and $rp$-process. During these calculations, intermediate states were sampled at randomly selected times, and the corresponding $X$, $T$, and $\rho$ values were used as inputs for Type A, whereas randomly rescaled $X$ values were used as inputs for Type B. The remaining inputs for Type A and II, as well as all inputs for Type C, were randomly sampled within the ranges in Table~\ref{tab_range}. Each input set was then evolved again with $\texttt{PRISM}$ at fixed $T$ and $\rho$ until the target time step $\Delta t$ was reached. The abundance at the end of the evolution was used as the label. The corresponding reaction rates and $f(Y_i; T, \rho)$ were also calculated during $\texttt{PRISM}$ runs, and nuclear data was assembled from standard sources; e.g. decay properties from \cite{M.Mumpower2025}, separation energies from FRDM \cite[]{P.Moller2016}, as in Ref.~\cite{Y.Zhu2021}.

\subsection{\label{sec_data_b}Data Preprocessing}

One challenge in this task is the large dynamic range of the data, which spans many orders of magnitude. We therefore applied a $\log_{10}$ transformation to $X$, $T$, $\rho$, $\Delta t$, and reaction rates. As the net flux term, $f(Y_i; T, \rho)$, can take both positive and negative values, it was transformed using a signed logarithm, $\text{sign}(f) \cdot \log_{10} (|f|)$. 

Instead of using the abundance at the next time step, $X_{t+\Delta t}$, as the label, we used the abundance change $\Delta X$. This choice is more appropriate because the physically relevant update over a time step is determined by $\Delta X$, not by the absolute magnitude of $X_{t+\Delta t}$. In many cases, $\Delta X$ is much smaller than $X_{t+\Delta t}$, so a small relative error in predicting $X_{t+\Delta t}$ can correspond to a very large relative error in $\Delta X$. 

\begin{figure}[t]
\centering
	\includegraphics[width=0.48\textwidth]{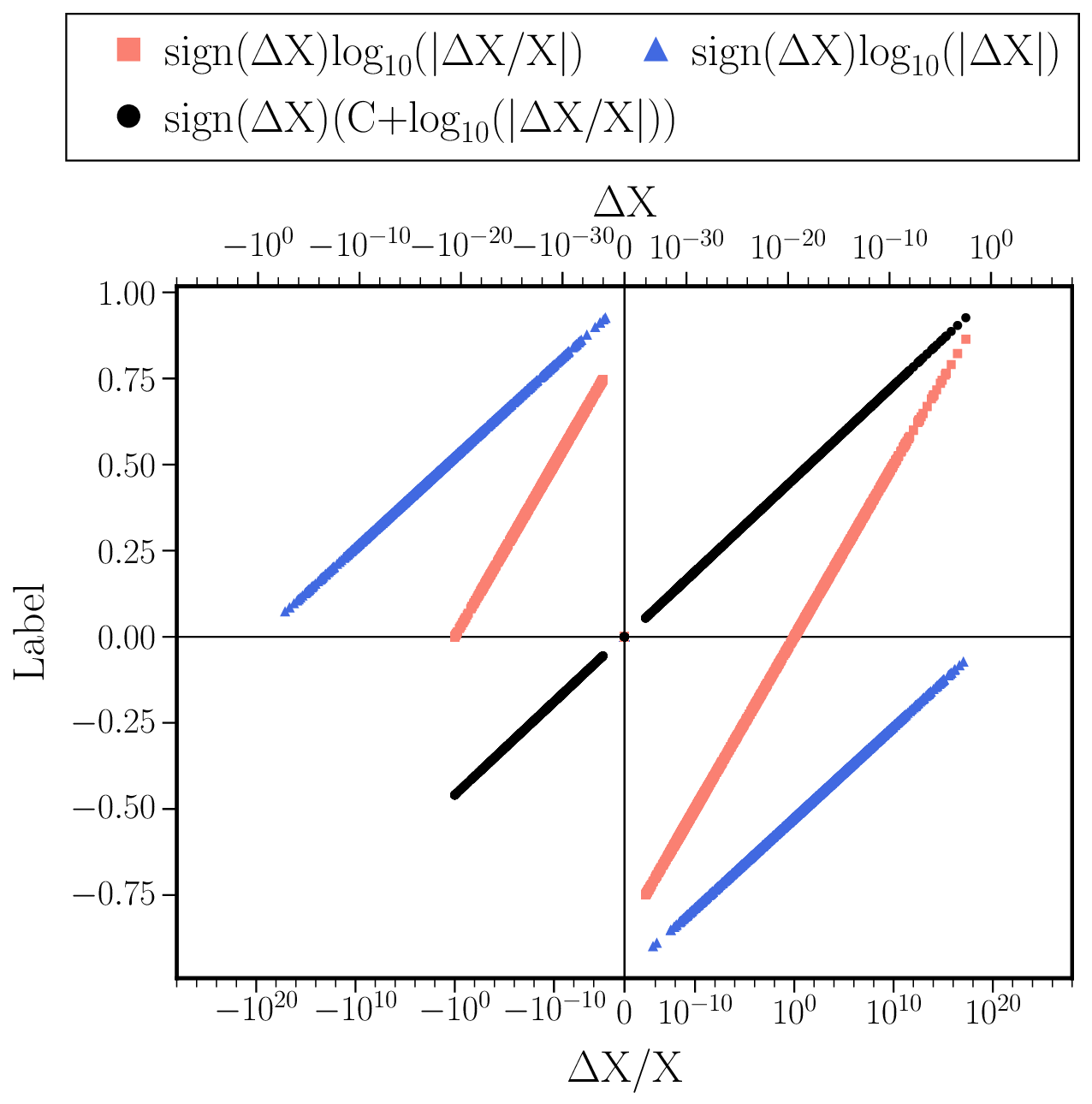}	 
\caption{\label{fig_delx}Transformed labels under different preprocessing schemes, shown as functions of $\Delta X$ ($\Delta X/X$). $\text{sign}(\Delta X) \cdot \log_{10}(|\Delta X|)$ is referenced to the upper x-axis ($\Delta X$), whereas the others are referenced to the lower x-axis ($\Delta X/X$). $\Delta X/X$ spans a narrower dynamic range than $\Delta X$. $\text{sign}(\Delta X) \cdot (C + \log_{10}(|\Delta X/X|))$ provides a more compact range while preserving the correct magnitude ordering and avoiding a disconnection at zero.}
\end{figure}

Both in the original reaction network solver and surrogate solvers in this study, the updated abundance is obtained as $X_{t+\Delta t} = X_{t} + \Delta X$. Because this update is performed in finite-precision arithmetic, increments smaller than roughly the machine epsilon $\epsilon_{\text{m}}$ relative to the existing abundance $X_{t}$ can be lost in the summation. For this reason, $\texttt{float64}$ ($\epsilon_{\text{m}}$=$2^{-52} \approx 2.2\times10^{-16}$) is normally required for a network solver. This numerical consideration also affects the choice of label preprocessing. Figure~\ref{fig_delx} compares possible preprocessing schemes for $\Delta X$. A direct signed log transformation of $\Delta X$ may not be ideal, because it leads to a broad target range. Since $X$ was clipped at $10^{-20}$, $\Delta X$ can be as small as $\sim$10$^{-36}$, given $\epsilon_{\text{m}}$ of $\texttt{float64}$. For this reason, we found it more effective to use the signed logarithm of the relative abundance change, $\Delta X/X$, for the label: $\text{sign}(\Delta X) \cdot \log_{10} (|\Delta X/X|)$. In addition, values with magnitude below $10^{-15}$ were clipped at zeros, since abundance relative changes near $\epsilon_{\text{m}}$ are dominated by numerical round-off rather than physically meaningful evolution. 

However, a few issues arise when applying a log transformation to quantities such as $\Delta X / X$ and $f(Y_i; T, \rho)$, which can be positive or negative and whose magnitudes are often smaller than 1. First, the logarithm gives a negative value even when the original value is positive, if it is smaller than 1. Second, for values smaller than 1, smaller original magnitudes are mapped to larger absolute values after the log transformation. For example, $\log_{10}(10^{-10})=-10$, while $\log_{10}(1)=0$. As a result, a naive signed log transformation, $\text{sign}(\Delta X) \cdot \log_{10}(|\Delta X/X|)$, reverses the magnitude ordering in the transformed space as shown in Figure~\ref{fig_delx}. To avoid this, we added a constant after the log transformation so that the log values become positive while preserving the original magnitude ordering before applying the sign: $\text{sign}(\Delta X) \cdot (C + \log_{10}(|\Delta X/X|))$. Here, $C$ was 17 which is about the maximum absolute magnitude of $\log_{10}(|\Delta X/X|)$ when $|\Delta X/X|$ is lower than 1. This effect is illustrated clearly in Figure~\ref{fig_delx}. 


We note that nearly all modern deep learning software and hardware are optimized for single-precision and lower precision formats, with $\texttt{float32}$ mostly used as the default precision, rather than $\texttt{float64}$. Double-precision computations are often much slower and less practical in standard deep learning frameworks. In addition, the larger memory cost of $\texttt{float64}$ further limits its efficiency. Nevertheless, the use of $\texttt{float32}$ did not discard very small abundance changes, because the labels were represented in logarithmic space. After the network prediction, the output was transformed back and cast to $\texttt{float64}$ before evaluating $X_{t+\Delta t} = X_{t}+\Delta X$. This approach allowed efficient single-precision inference while preserving small $\Delta X$ values reliably.

\section{\label{sec_net}Network Design}

\subsection{\label{sec_net_gnn}Graph Neural Network}

\begin{figure*}[t]
\centering
	\includegraphics[width=0.9\textwidth]{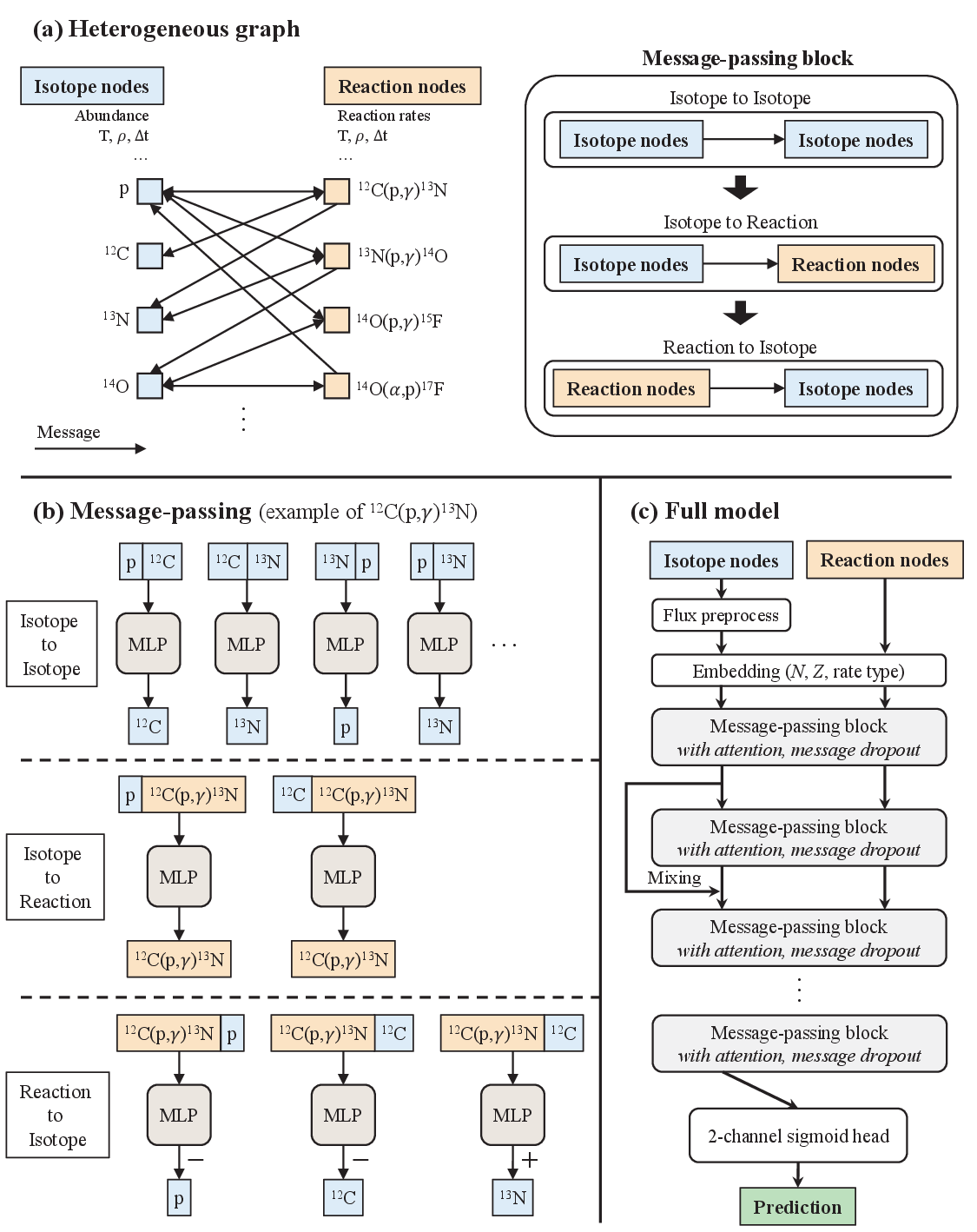}	 
\caption{\label{fig_gnn}Overview of NuGNN architecture. (a) Illustration of the graph composed of two different types of nodes: isotope and reaction nodes. (b) Examples of message-passing where each node exchanges messages with neighboring nodes connected through nuclear reactions. (c) Full model architecture, including flux preprocessing, embedding, message-passing blocks, and the two-channel sigmoid output head. See the text for details.}
\end{figure*}

\subsubsection{\label{sec_net_gnn1}Graph representation of reaction network}

From a structural point of view, a nuclear reaction network can naturally be represented as a graph. Isotopes and reactions form interconnected nodes and edges that encode how abundance flows through the network. This graph-based structure motivates the use of graph neural networks as surrogate models. A graph neural network (GNN) is a neural network designed to process data represented as a graph, where objects are expressed as nodes and their relationships are expressed as edges \cite[]{F.Scarselli2009}. Unlike standard feed-forward networks, a GNN updates the representation of each node by aggregating information from its neighboring nodes through the graph connections \cite[]{J.Gilmer2017}. By repeating this process known as `message-passing' over multiple layers, the model can learn both local interactions and global structure in the graph.

\subsubsection{\label{sec_net_gnn2}Isotope and reaction nodes}

Figure~\ref{fig_gnn} shows a schematic of the GNN in this study (NuGNN). We represented the nuclear reaction network as a heterogeneous graph, or bipartite graph, composed of isotope nodes and reaction nodes. Each isotope node is assigned features consisting of the mass fractions of the isotope, net flux values, reaction rates mapped onto the isotope grid, separation energies of one and two protons, $T$, $\rho$, $\Delta t$, and learned embeddings of neutron and proton numbers. Each reaction node is assigned the corresponding reaction rate, $T$, $\rho$, $\Delta t$, and learned embedding of reaction type---$(p,\gamma)$, $(\gamma,p)$, $(\alpha,\gamma)$, $(\gamma,\alpha)$, $(p,\alpha)$, $(\alpha,p)$, and $\beta^+$.

\subsubsection{\label{sec_net_gnn3}Flux preprocessor}

Before message-passing, the net flux in the isotope features was refined by a dedicated preprocessing block. This was motivated by the fact that the net flux often provides the correct sign and approximate relative magnitude of the abundance changes, but still contains inaccurate values that can confuse the network if used directly. The model therefore learns an input-dependent correction to the flux channel using the isotope features, as well as from the magnitude and sign of the flux. In this way, the network can preserve useful flux information while suppressing unreliable contributions before graph propagation begins.

\subsubsection{\label{sec_net_gnn4}Message-passing}

The graph contains three types of connections that reflect the structure of the reaction network, as shown in Figure~\ref{fig_gnn} (a): reactant isotopes are connected to the corresponding reactions (Isotope to Reaction), each reaction is connected to its reactant and product isotopes (Reaction to Isotope), and direct isotope-to-isotope connections are included to represent couplings between nuclides involved in the same reaction system (Isotope to Isotope). 

The network updates this graph through multiple message-passing layers. Figure~\ref{fig_gnn} (b) shows an example of message-passing in this model. In each layer, messages are passed directly between isotopes, from isotopes to reactions, and from reactions to isotopes. The Isotope to Isotope step propagates information between isotopes coupled through reactions. The Isotope to Reaction step collects information from the reactant isotopes into the corresponding reaction node. The Reaction to Isotope step distributes the updated reaction information back to the reactant and product isotopes. In this step, signed coefficients are included so that the model preserves whether a reaction contributes negatively (reactants) or positively (products) to the abundance change of a given isotope. These updates are performed separately for each reaction type so that different reaction classes are treated with different learned neural network layers.

\subsubsection{\label{sec_net_gnn5}Attention, dropout, and mixing}

Within each message-passing step, attention is used to weight the importance of messages coming from neighboring nodes before aggregation \cite[]{P.Velickovic2018}, along with sigmoid gates recently suggested by \cite{Z.Qiu2025}. Message dropout is also applied during message-passing to make the learned node representations more robust to the presence or absence of individual connections \cite[]{X.Wang2019}. 

Because abundance changes typically scale with $T$, $\rho$, and $\Delta t$, we allowed the model to adapt how much information from an early layer is retained or suppressed according to these conditions. Specifically, the early and later isotope representations were mixed using coefficients learned from $T$, $\rho$, and $\Delta t$, enabling the network to control the effective message-passing depth to the expected scale of abundance evolution.

\subsubsection{\label{sec_net_gnn6}Output head and architectural details}

In the final stage, the isotope representation of each node is mapped to two output channels. After applying a sigmoid activation to both channels, the second channel is subtracted from the first, yielding the final prediction. Since each channel lies in $[0,1]$, their difference is bounded in $[-1,1]$, which covers the range of the preprocessed label shown in Figure~\ref{fig_delx}. 

The NuGNN consists of five message-passing blocks. Specifically, each block contains separate neural network layers for the three message pathways---Isotope to Isotope, Isotope to Reaction, and Reaction to Isotope---and these layers are further separated by reaction type. Therefore, each reaction type has its own learned message and update functions, allowing the model to treat different nuclear processes with different parameters. Each of these message and update functions is implemented as two fully connected layers with a hidden width of 64 channels and activation function as Leaky ReLU \cite[]{A.Maas2013}.

\subsection{\label{sec_net_unet}Res-U-Net}

As a model for comparison, we also employed a residual U-Net (Res-U-Net). Unlike the GNN, which explicitly follows the connectivity of the reaction network, the Res-U-Net treats the preprocessed nuclear quantities as multi-channel two-dimensional maps (i.e., images) on the $(N,Z)$ plane. In this representation, each channel corresponds to one of the input quantities described in Section~\ref{sec_data}, arranged on the nuclear chart. This allows the model to learn abundance evolution through spatial patterns on the nuclear map.

The architecture follows the standard encoder-decoder structure of U-Net \cite[]{O.Ronneberger2015}. In the encoder, a sequence of convolutional blocks progressively reduces the spatial resolution while increasing the number of feature channels, enabling the network to extract abstract features over a wider region of the nuclear chart. After passing through the bottleneck, the decoder gradually restores the spatial resolution by upsampling the feature maps. At each decoder stage, the upsampled feature map is concatenated with the corresponding encoder feature map through a skip connection. These skip connections allow the model to preserve fine local information while also incorporating the broader contextual information learned at deeper levels. Each convolutional block in the encoder or decoder is implemented as a residual block, where the block input is added to the output features \cite[]{K.He2016}. This residual design improves optimization and stabilizes training in a deep convolutional network. 

In addition to the two-dimensional map inputs, the model also takes global state quantities ($T$, $\rho$, and $\Delta t$) as a separate input. These quantities are processed by a feature-wise linear modulation (FiLM) network, which generates channel-wise scaling and bias values \cite[]{E.Perez2018}. These modulation values are applied to the convolutional feature maps, allowing the extracted spatial features to adapt to the thermodynamic state and time step size. In this way, the Res-U-Net combines local spatial information from the nuclear chart with global physical information that controls the overall evolution.

The network consists of three encoder blocks, a bottleneck block, and three decoder blocks, with channel sizes progressing as 64, 128, 256, 512, 256, 128, and 64. Each block of encoder and decoder has eight convolutional layers with batch normalization \cite[]{S.Ioffe2015}, and the bottleneck block has two of them. The final output layer follows the same two-channel method described in Section~\ref{sec_net_gnn6}. Leaky ReLU is also used as the activation function throughout the network.

\subsection{\label{sec_net_fnn}Fully connected neural network}

As another reference model, we employed a fully connected neural network (FNN), as done in previous studies \cite[]{D.Fan2022, A.Grichener2025, X.Zhang2025}. Unlike the GNN, which explicitly uses the connectivity of the nuclear reaction network, and unlike the Res-U-Net, which exploits spatial locality on the $(N,Z)$ map, this model treats each isotope rather independently at the architectural level. For each isotope, the input consists of the same preprocessed quantities used in the other models, arranged as a one-dimensional feature vector rather than as graph nodes or image-like maps. The network also includes residual-style additions between hidden layers. The final output layer follows the same two-channel method described in Section~\ref{sec_net_gnn6}. It consists of five fully connected layers in total, with a hidden width of 1024 channels and Leaky ReLU as the activation function. 

\begin{figure}[t]
\centering
	\includegraphics[width=0.48\textwidth]{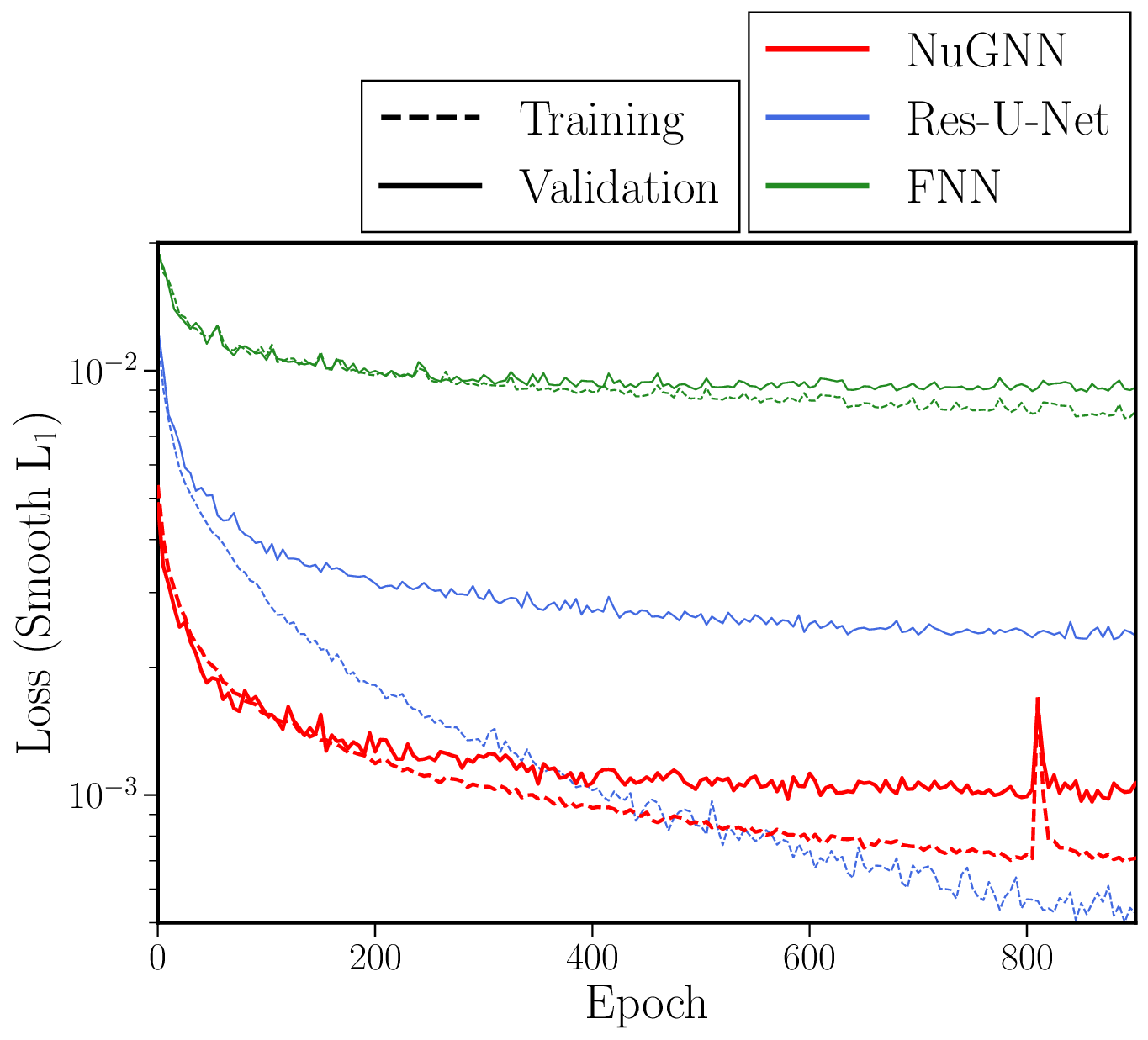}	 
\caption{\label{fig_curve}Learning curves for NuGNN, Res-U-Net, and FNN. The validation loss indicates that NuGNN outperforms the other two architectures. For clarity, only the epoch with the lowest RMS within each 5-epoch interval is shown.}
\end{figure}

\section{\label{sec_train}Training Method}

A total of 90,000 samples---30,000 samples for each data type mentioned in Section~\ref{sec_data_a}---were randomly shuffled and divided into training, validation, and test datasets with ratios of 0.8, 0.1, and 0.1, respectively. The model was trained by minimizing a Smooth $L_1$ loss, which behaves as the mean squared error ($L_2$) when the absolute error is smaller than $\beta_L$ and as the mean absolute error ($L_1$) otherwise, where $\beta_L = 0.001$ in this study \cite[]{R.Girshick2015}. This loss function is less sensitive to rare large deviations than the mean squared error, while still providing stable gradients near the target value. To emphasize physically important cases, we applied element- and data type-wise weights in the loss: the $p$ and $\alpha$ channels were given weights 10 times larger than those of the other isotopes, and Type C samples were assigned smaller weights than samples of the other types. This weighting was introduced to encourage more accurate predictions for species that are especially important for the nuclear evolution and to prevent the optimization from being dominated by less critical targets.

The mass fractions should satisfy $\sum_i X_i = 1$, or equivalently $\sum_i \Delta X_i = 0$.  Although the model can in principle learn this constraint implicitly from the data, it could also be enforced explicitly by introducing an additional physics-informed loss term or by designing the final activation function of the network to satisfy mass conservation, as done in previous studies \cite[]{D.Fan2022, A.Grichener2025, X.Zhang2025}. However, in this study, we found that such approaches made the training unstable because the target labels were preprocessed into logarithmic form (Section~\ref{sec_data_b}). Enforcing the explicit conservation constraint requires transforming the network outputs back from the signed log representation to $\Delta X$, after which their sum can be evaluated and used either in the loss function or in the output transformation. In practice, this inverse transformation made the optimization unstable and degraded the training performance. Still, we found that the model learned the conservation condition sufficiently well from the data alone even without an additional enforcing term.

We used the Adam optimizer for training \cite[]{Adam}. The learning rate was set to $10^{-3}$ for the first 1,000 epochs. After that, the best-performing checkpoint from those 1,000 epochs was further trained for 200 additional epochs with a learning rate of $10^{-4}$. This two-stage schedule allowed the model to refine its parameters more stably at a smaller learning rate.


The models were trained in $\texttt{PyTorch}$ \cite[]{A.Paszke2019} and then converted into $\texttt{NVIDIA}$ $\texttt{TensorRT}$ engines. $\texttt{TensorRT}$ is a software development kit for high-performance deep learning inference. It provides an optimized engine for low-latency, high-throughput inference on GPUs by performing optimizations such as efficient kernel selection and memory reuse. In this study, $\texttt{TensorRT}$ was used to accelerate the surrogate solver evaluation after training, with the aim of improving the practical applicability of the model in large-scale simulations. 

\begin{table}[t]
\centering
\caption{\label{tab_error}Mean absolute errors of each model on the test dataset. Both the total error and the error for non-zero mass fractions $X$ are shown. Each error is further reported separately for each data type. NuGNN yields errors approximately 3--10 times lower than those of the other two architectures.}
\begin{ruledtabular}
\begin{tabular}{c|cccc} 

All $X$ & NuGNN & Res-U-Net & FNN \\ \hline
Total Error & 0.037 & 0.11 & 0.56 \\
Type A & 0.0079 & 0.020 & 0.034 \\ 
Type B & 0.024 & 0.054 & 0.22 \\
Type C & 0.081 & 0.27 & 1.45 \\ \hline \hline

Non-zero $X$ & NuGNN & Res-U-Net & FNN \\ \hline
Total Error & 0.11 & 0.32 & 1.69 \\
Type A & 0.035 & 0.087 & 0.15 \\ 
Type B & 0.094 & 0.22 & 0.87 \\ 
Type C & 0.15 & 0.46 & 2.67 \\ 

\end{tabular}
\end{ruledtabular}
\end{table}

\begin{figure*}[t]
\centering
	\includegraphics[width=1\textwidth]{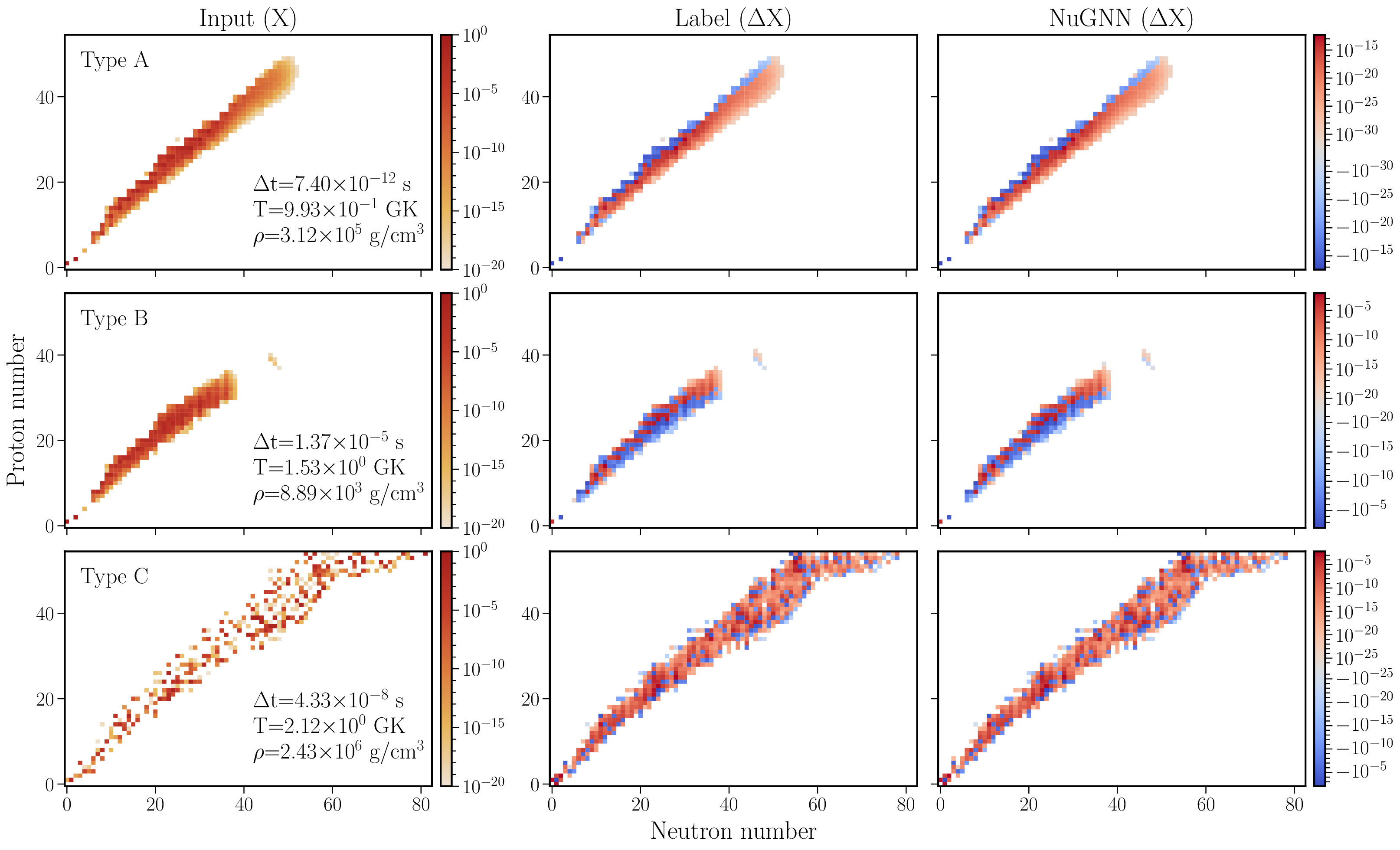}	 
\caption{\label{fig_examples}Examples of NuGNN predictions for the test dataset. NuGNN reproduces the true $\Delta X$ (Label) remarkably well, even though the data span tens of orders of magnitude in both the negative and positive directions.} 
\end{figure*}


\section{\label{sec_results}Results}

Figure~\ref{fig_curve} shows the learning curves of the NuGNN, Res-U-Net, and FNN on the training and validation datasets. The NuGNN consistently achieves lower errors than the other architectures throughout training. A quantitative comparison on the test dataset is given in Table~\ref{tab_error}, which reports the mean absolute error in the signed logarithm of relative abundance change introduced in Section~\ref{sec_data_b}:
\begin{align}
\text{Error} =& \frac{1}{N}\sum_i^N \Bigl| \text{sign}(\Delta X^\text{t}_i)\cdot \left(C+\log_{10}\left|\Delta X^\text{t}_i/X^\text{t}_i\right|\right) \notag\\
&\qquad - \text{sign}(\Delta X_i)\cdot \left(C+\log_{10}\left|\Delta X_i/X^\text{t}_i\right|\right) \Bigr| \label{eq_error1} \\
=& \frac{1}{N}\sum_i^N \left| \log_{10}\left|\Delta X_i^\text{t}/\Delta X_i\right| \right|,  \notag\\
& \text{if } \text{sign}(\Delta X^\text{t}_i)=\text{sign}(\Delta X_i),\label{eq_error2}
\end{align}
where $N$ is the number of data samples, $i$ is the sample index, and the superscript `t' denotes the true value. When the predicted and true signs agree, the error reduces to the difference between the logarithms of the true and predicted magnitudes. When the predicted sign is incorrect, the two terms add rather than cancel, so the error becomes larger and sign disagreement additionally contributes to the error. The accuracy on the sign was $\sim$97$\%$ for the case of NuGNN. 

\begin{figure*}[t]
\centering
	\includegraphics[width=1\textwidth]{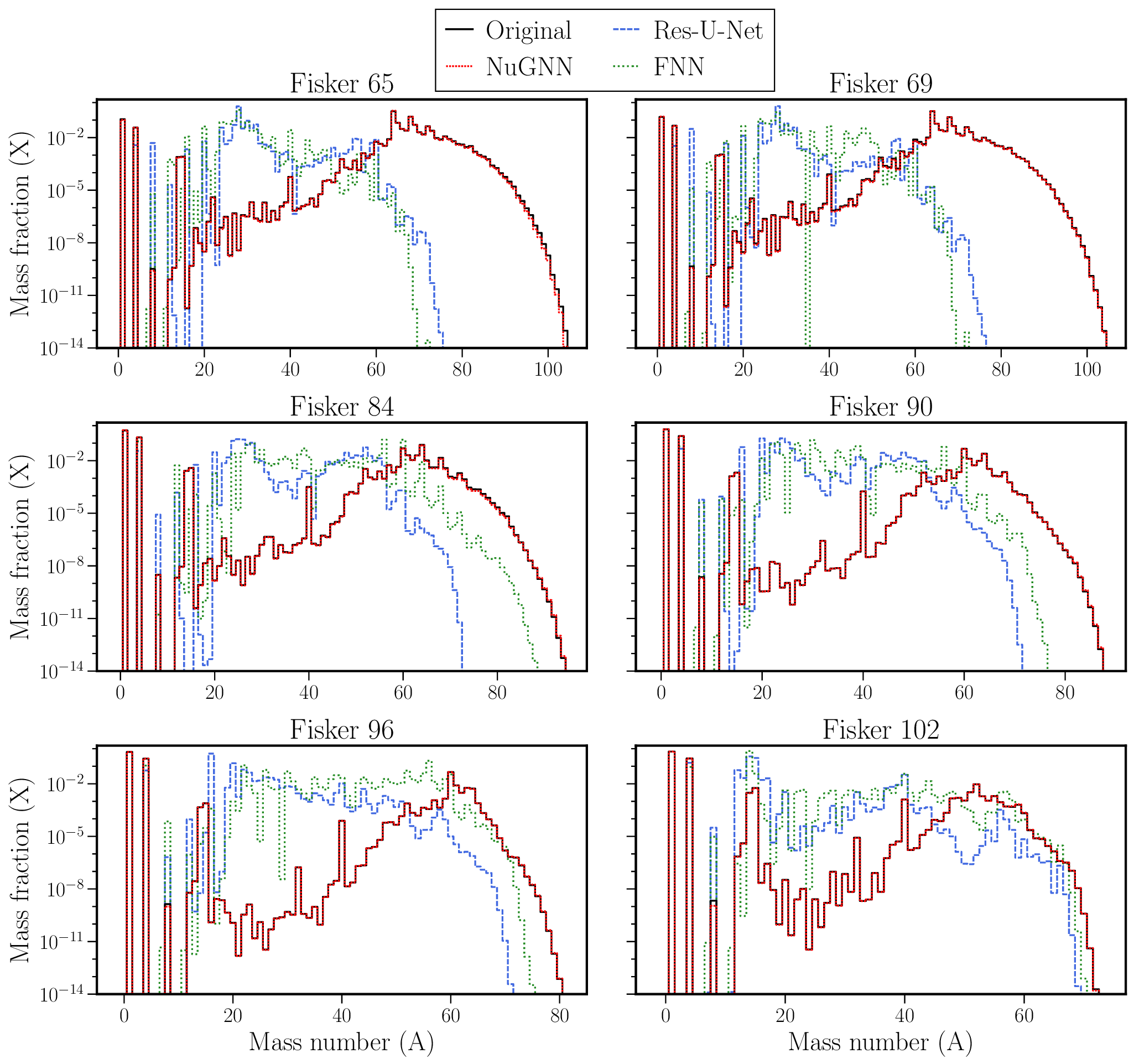}	 
\caption{\label{fig_abundance}Final abundances obtained with different solvers for different thermodynamic trajectories from \cite{J.Fisker2005,J.Fisker2008} which were not used for training. `Original' denotes the final abundances evolved using the original solver. The title of each subplot indicates the zone number of the simulations of \cite{J.Fisker2005,J.Fisker2008}. While Res-U-Net and FNN were unable to accurately reproduce the final abundance patterns, NuGNN produced highly accurate final abundance patterns for all zones.}
\end{figure*}

Table~\ref{tab_error} reports the errors separately for the three types of data introduced in Section~\ref{sec_data_a}, considering both all mass fractions and non-zero mass fractions only. The latter distinction is useful because accurately predicting non-zero mass fractions is generally more difficult than predicting zero values. Because the error metric is defined on a $\log_{10}$ scale (Equations~\ref{eq_error1} and \ref{eq_error2}), it approximately reflects the discrepancy in the order of magnitude of $\Delta X$ and $\Delta X/X$. For Type A, which corresponds to physically realistic data, the errors can be roughly interpreted in linear space as 1.8$\%$ and 8.4$\%$ in $\Delta X$ for NuGNN, 4.7$\%$ and 22.2$\%$ for Res-U-Net, and 8.1$\%$ and 41.3$\%$ for FNN, for all and non-zero mass fractions, respectively. This interpretation is only approximate, because it assumes that the predicted signs are correct for all samples, and because a mean absolute error defined in log space may not be converted exactly into a mean absolute error in linear space.

Figure~\ref{fig_examples} shows examples of predictions ($\Delta X$) on $(N,Z)$ plane for different data types and for different values of $T$, $\rho$, and $\Delta t$. Overall, the predictions are in excellent agreement with the labels. As discussed in Section~\ref{sec_data_a}, Type A and Type B data exhibit realistic abundance patterns, whereas Type C data are generated from random abundance patterns, as can be seen in the figure. This characteristic makes Type C cases more challenging for the model to predict accurately, which is also reflected in Table~\ref{tab_error}. Still, Figure~\ref{fig_examples} shows that NuGNN predicts Type C cases accurately, reproducing the signs and magnitudes correctly in most regions.

For practical evaluation, we replaced the original solver in the network evolution code ($\texttt{PRISM}$) with NuGNN, Res-U-Net, and FNN solvers. We performed calculations using six different thermodynamic profiles of X-ray bursts for $\sim$300 seconds from the study of \cite{J.Fisker2005,J.Fisker2008}, which were not included in the training data. Figure~\ref{fig_abundance} presents the predicted final abundances for each profile. Res-U-Net and FNN failed to reproduce the abundance patterns at all. In particular, both models overly deplete protons and alphas---the species that serve as the fuel for most of the reactions. Additionally, in contrast to NuGNN, Res-U-Net and FNN could not achieve sufficient single-step accuracy (See Table~\ref{tab_error}); consequently, errors accumulate over many time steps and the abundance evolution drifts toward an incorrect solution. Nucleosynthesis of these models proceeds only up to about $A\sim80$, and their predicted abundances differ from those of the original solver by several orders of magnitude. In contrast, NuGNN successfully reproduced the final abundances of the X-ray bursts even for proton and $\alpha$, with deviations from the original solver remaining mostly below $\sim$15$\%$.

\begin{figure}[t]
\centering
	\includegraphics[width=0.48\textwidth]{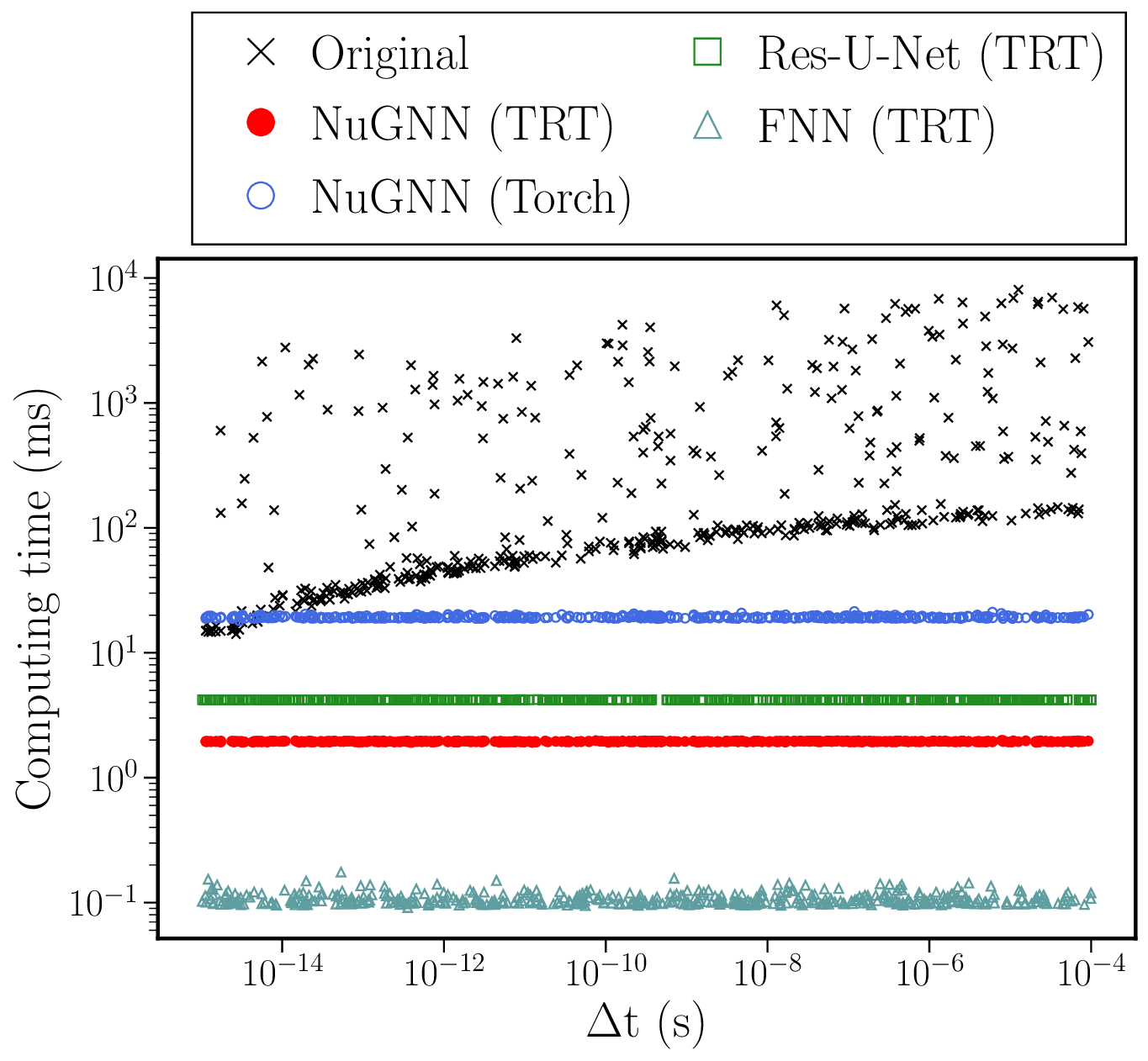}	 
\caption{\label{fig_time}Computing (inference) times on test data samples for the original and different surrogate solvers in this study. `Torch' denotes the inference time using the trained $\texttt{PyTorch}$ model, whereas `TRT' denotes the $\texttt{TensorRT}$ engine built from the trained model.}
\end{figure}

Figure~\ref{fig_time} shows the computing times we measured on test data samples for different solvers. We used Intel Xeon Platinum 8360Y for the original solver and NVIDIA RTX 5090 for the surrogates. The inference times of the surrogates do not depend on the input data as shown in the figure. Graph neural networks are generally slower than conventional neural networks because they rely on computationally expensive operations that are less favorable for modern hardware acceleration, particularly the scatter and gather operations required to pass messages between nodes. This is reflected in the figure by the data points labeled `NuGNN (Torch)', which shows inference times of $\sim$19.3 ms. However, we found that converting the $\texttt{PyTorch}$ model into a $\texttt{TensorRT}$ engine substantially improved the inference speed. The resulting inference times were 1.9, 4.2, and 0.1 ms for the $\texttt{TensorRT}$ implementations of NuGNN, Res-U-Net, and FNN, respectively. Res-U-Net remained relatively slow because it processes heavy image-like inputs through convolutional layers, whereas FNN was the fastest due to its simple feed-forward structure. 


\section{\label{sec_future}Conclusions and Future Work}

In this study, we used a deep learning approach to develop a surrogate solver for a 690-isotope nuclear reaction network under Type I X-ray burst conditions. By explicitly reflecting the structure of the reaction network through isotope and reaction nodes, our graph neural network NuGNN had very good performance on our tests of abundance predictions and network evolution, while the Res-U-Net and fully connected neural network performed substantially worse on these tests and failed to produce viable abundance patterns. We also showed that the trained model can be converted into a $\texttt{TensorRT}$ engine, providing a significant improvement in inference speed and demonstrating its potential for practical use in large-scale simulations. Overall, these results show that graph neural networks are a promising approach for building fast and accurate surrogate solvers for large nuclear reaction networks. 

There are still several aspects that require further improvement in the future. The models were trained only on mass fractions larger than $10^{-20}$, while values below this threshold were treated as zero in the data preprocessing (see Section~\ref{sec_data_a}). As a result, the models are not expected to predict abundances smaller than $10^{-20}$ reliably. In addition, because the prediction can occasionally become less accurate for a given time step, a retry procedure was still beneficial in the network evolution code. Specifically, if the sum of the predicted $\Delta X$ deviated significantly from zero, the evolution step was retried with a smaller $\Delta t$, for which the prediction task is easier for the model. 

Further developments of the model and its implementation are possible in several ways. The extrapolation of deep learning models is generally not reliable \cite[]{C.Kim2026}, and broad training data coverage is therefore essential for robust generalization. In this context, uncertainty quantification can be useful for identifying cases in which the model may fail, particularly for out-of-distribution inputs or extrapolation regimes. Such uncertainty estimates can be performed using Bayesian deep learning such as Monte Carlo dropout or deep ensembles \cite[]{Y.Gal2016, B.Lakshminarayanan2017, C.Kim2024}. They could also be integrated into the retry procedure, for example by using large predictive uncertainty as a criterion for reducing the time step. Additionally, the graph neural network architecture is flexible and could be further improved through changes in components such as the attention mechanism, message-passing scheme, aggregation function, and so on. It may also be beneficial to incorporate the importance of key fuel species, such as protons and $\alpha$ particles, more directly at the architectural level. 

The present framework is also expected to benefit from future advances in GPU hardware and software. Such advances will provide greater computational speed and memory capacity, while enabling the use of larger models and more extensive training data. First, a wider range of training data samples will become more feasible. This includes lower mass fractions even below $10^{-20}$ and broader ranges of temperature, density, and time step size, thereby covering more general as well as more specialized astrophysical conditions. Second, larger models and more training data may further improve predictive performance, which could reduce the uncertainty of the model predictions. Third, although GNNs are typically slower than the standard network architectures, future hardware advances are expected to improve the speed and memory efficiency of GNNs. Further gains will also depend on software improvements, which can better handle irregular memory access and aggregation over sparse graph connections. Finally, these advances can enable the development of surrogate models for larger reaction networks involving heavier nucleosynthesis, such as the $r$-process.

\begin{acknowledgments}

This work was supported by the National Research Foundation of Korea (NRF) grants funded by the Korea government (MSIT) (Grant No. RS-2024-00338255). Computational works for this research were performed partly on the data analysis hub, Olaf in the IBS Research Solution Center. 

\end{acknowledgments}


\end{document}